\documentstyle[titlepage,12pt]{article}

\begin{document}
\hfill KL-TH 96/6\\

\thispagestyle{empty}
\begin{center}

{\large \bf APPLICATION OF INSTANTONS:\\
QUENCHING OF MACROSCOPIC QUANTUM COHERENCE
AND
MACROSCOPIC FERMI--PARTICLE
CONFIGURATIONS
\\[14mm]}
{\large J.--Q. Liang}\\
{\it Department of Physics, University of Kaiserslautern, D--67653 Kaiserslautern,
Germany,\\
Institute of Theoretical Physics, Shanxi University, Taiyuan, Shanxi 030 006, P.R.China,
Institute of Physics, Academia Sinica, Beijing 100 080, P.R. China}\\[1cm]

{\large H.J.W. M\"uller-Kirsten and Jian-Ge Zhou}\\
{\it Department of Physics, University of Kaiserslautern, D--67653 Kaiserslautern, Germany}\\
[1cm]

{\bf Abstract}\\
\end{center}
Starting from the coherent state representation of the evolution operator
with the help of the path--integral, we derive a formula for the low--lying
levels $E = \epsilon_0 - 2\triangle\epsilon\cos (s+\xi)\pi$ of a quantum spin system.
The quenching of macroscopic quantum coherence is understood as the vanishing
of $\cos (s+\xi)\pi$ in disagreement with the suppression of tunneling
(i.e. $\triangle\epsilon = 0$) as claimed in the literature.  A new configuration called
the macroscopic Fermi--particle is suggested by the character of its wave
function.  The tunneling rate ($\frac {2\triangle\epsilon}{\pi}$) does not vanish,
not for integer spin $s$ nor for a half--integer value of $s$, and is 
calculated explicitly (for the position dependent mass) up to the
one--loop approximation.\\
%\end

\newpage
The question whether or not superpositions of macroscopically
distinct quantum states exist in nature has attracted considerable attention in recent
years owing mainly to the development of technology in mesoscopic physics. Thus
Leggett et al.\cite {one}
predicted that the most intriguing quantum effect which could take place
on the macroscopic scale is quantum tunneling\cite {one}.  Macroscopic
magnetisation tunneling is a subject which is being investigated extensively
 and is of growing interest \cite{two, three}.  In discussing this, it is essential
to distinguish between macroscopic quantum tunneling (MQT) and macroscopic
quantum coherence (MQC).  In the case of MQC, tunneling between neighboring 
degenerate (perturbation theory) vacua is dominated by the instanton configuration with nonzero
topological charge and leads to a level splitting which (apart from a 
factor of $\hbar$ ) is seen to be the frequency of oscillation between, for 
example, the easy directions (defined by the vacua) of the ferromagnetic particle.  The tunneling
removes the degeneracy of the ground states, and the true ground state is
a superposition of the degenerate ground states, for instance, the two easy 
directions of the ferromagnetic particle which are macroscopically distinguishable. 
For the case of MQT the tunneling is dominated by the socalled bounce configuration \cite{four}
with zero topological charge and leads to the decay of metastable states.
The quantum tunneling of magnetization in small ferromagnetic particles has been well
established within the leading order approximation of the instanton method for 
various models \cite{five}, using the classical Hamiltonian.
Recently an interesting phenomenon known as topological quenching of MQC has been
studied extensively \cite{six, seven, eight, nine}. The quenching of MQC there was
analysed as the suppression of tunneling (i.e. the vanishing of the level splitting
$\triangle\epsilon$ as given by our eqs.(14,15) below which gives its relation to the
transition amplitude ) by destructive interference of Euclidean paths in the barrier \cite{six}.
In the case of MQC the Hamiltonian has a periodic nature,
that is, there is an infinite number of degenerate vacua by successive $2\pi$ extension.
However, tunneling in the
ferromagnetic particle has only been considered in the literature so far
as that for the double--well potential,  
like the $\phi^4$ potential in field theory, where the side barriers are
infinitely high.  We argue here that the method of calculating the tunneling
effect for a periodic potential \cite {ten} ought to be used in dealing with MQC,
regarding the periodic potential as a one--dimensional super--lattice.  The
tunneling through one barrier leads to the level splitting which extends formally to an
energy ``band'' by translation symmetry (the rotation symmetry in the present
case, $V(\phi + n\pi) = V(\phi))$. The energy ``band'' is formally the same as that
of a one--dimensional tight--binding model in solid state physics.  If one proceeds like this, an
interesting observation emerges.  The energy ``band'' for half--integer spin
$s$ is quenched to a single degenerate level (doubly degenerate in the present case)
in agreement with Kramer's theorem
which says that for half--integer spin s the degeneracy
cannot be removed in the absence of the magnetic field. 
The quenching of MQC is then understood as due to the quenching of the
energy band, and is the result of the destructive interference of Bloch waves.
From a theoretical point of view it may be interesting to note
that the wave function of the giant particle
is a superposition of macroscopically distinct states with
a phase resulting from the Wess--Zumino term. This phase changes its sign  under a $2\pi$--rotation.  
The ferromagnetic particle thus considered can therefore be regarded as a
macroscopic Fermi--particle configuration. To see the difference of the two
formalisms explicitly we start in the following from the spin coherent state path integral.

The Hamiltonian operator $\hat{H}$ of the ferromagnetic particle we consider is
\begin{equation}
\hat{H} = K_1  \hat{S_z}^2 + K_2 \hat{S_y}^2
\end{equation}
where $\hat{S_i}, i=x, y, z,$ are spin operators with the usual commutation relation
$[\hat{S_i}, \hat{S_j}] = i \epsilon_{ijk}\hat{S_k}$ (using natural units throughout) and $K_1> K_2>0$.
This model describes $XOY$--easy plane anisotropy with an easy axis along the x direction
\cite{five}.  We begin with the evaluation of the matrix element of the
evolution operator in the spin coherent-state representation by means of the
coherent-state path-integral, i.e.
\begin{equation}
<{\bf n}_f|e^{-2i\hat{H}T}|{\bf n}_i> = \int \{\Pi^{M-1}_{k=1} d\mu({\bf n}_k)\} \{\Pi^M_{k=1}
<{\bf n}_k|:e^{-i\epsilon\hat{H}}:|{\bf n}_{k-1}>\}
\end{equation}
Here we define $|{\bf n}_M> = |{\bf n}_f>, |{\bf n}_0> = |{\bf n}_i>$ and $t_f-t_i = 2T,
\epsilon = \frac {2T}{M}$ respectively.  Further $|{\bf n}>$ denotes the spin coherent--state 
generated from the reference spin eigenstate $|s,s>$ such that
\begin{equation}
|{\bf n}> = e^{\zeta^*\hat{S}_- - \zeta\hat{S}_+}|s,s>,
\end{equation}
where $\zeta = \frac{\theta}{2}e^{-i\phi}$ with polar angle $\theta$ and azimuthal angle $\phi$
of the unit vector ${\bf n} = (\sin\theta \cos\phi,\newline \sin\theta\sin\phi, \cos\theta)$. The
measure of integration is defined by \cite{eleven}
\begin{equation}
d\mu({\bf n}_k) = \frac{2s+1}{4\pi} d{\bf n}_k,\;  d{\bf n}_k = \sin\theta_k d\theta_k d\phi_k
\end{equation}  
The normal ordered infinitesimal evolution operator is  
\begin{equation}
:exp\{-i\epsilon\hat{H}\}: = exp\{i\epsilon\frac{K_2}{4}\hat{S}^2_-\}exp\{-i\epsilon[\frac{K_2}{2}\hat{S}^2
+(K_1-\frac{K_2}{2})\hat{S}^2_z]\}exp\{i\epsilon\frac{K_2}{4}\hat{S}^2_+\}
\end{equation}
The element in eq. (2) can be evaluated by regarding the coherent state formally as an
``eigenstate'' of operators $\hat{S}^2, \hat{S}_z$ and $\hat{S}_+$
with ``eigenvalues'' $s(s+1), s\cos\theta$ and $se^{i\phi}\sin\theta$ respectively, the
operators in eq.(2) then being replaced by their corresponding ``eigenvalues'' and the
coherent--state path--integral in eq.(2) is {\it formally} written as the path integral in 
phase space for large s, i.e.
\begin{equation}
e^{-i(\phi_f-\phi_i)s}\int{\cal D}{\phi}{\cal D}{p}e^{i\int^{t_f}_{t_i}{\cal L}(\phi,p)dt}
\end{equation}
with canonical variables $\phi$ and $p = s\cos\theta$, where
\begin{equation}
{\cal L} = \dot{\phi}p - H(\phi, p)
\end{equation}
is the phase space (or first order) Lagrangian, and the Hamiltonian
\begin{equation}
H = \frac {p^2}{2m(\phi)} + V(\phi)
\end{equation}
is just the classical Hamiltonian as in refs. \cite{five, twelve}. In $H$ the mass
\begin{displaymath}
m(\phi) = \frac{1}{2K_1(1 - \lambda\sin^2\phi)}
\end{displaymath}
is position dependent (reflecting the fact that there is a velocity dependent force) and
$V(\phi) = K_2s^2\sin^2\phi$ is the potential, and $ \lambda = \frac {K_2}{K_1} $
is a dimensionless parameter.
The position dependent mass also implies that the original space on which the 
continuous classical mechanics is defined for the spin system (see the following)
has an intrinsic curvature \cite{thirteen}.
The position dependent kinetic term corresponds to
a velocity dependent potential, and this requires that one start from the
phase space path--integral as pointed out long ago \cite{fourteen}.
The phase of the prefactor $ e^{-is(\phi_f-\phi_i)}$
which originated from the inner product of coherent states \cite{eleven, fifteen}, i.e.
\begin{displaymath}
\Pi^M_{k=1} <{\bf n}_k|{\bf n}_{k-1}>
\end{displaymath}
can be put into
the Lagrangian, i.e. the angle difference $\phi_f - \phi_i$ is written as
$ \phi_f - \phi_i =\int^{t_f}_{t_i}\dot{\phi}dt$, 
and has been identified formally as a Wess--Zumino term \cite{six, fifteen}. 
This phase factor,  which depends only
on the difference of initial and final azimuthal angles of the states $|{\bf n}_i> $ and $|{\bf n}_f>$ 
will be seen in the following to be just the phase difference of wave functions
in the coherent state representation.
We remark here that one cannot add $2\pi n$ (as in eq. (4) of ref.\cite {six}) into the
angle difference since the equality with the original phase factor $e^{-i(\phi_f -\phi_i)s}$
would be violated when $s$ is a half-integer.  This may be the technical point in ref.\cite {six}
which leads to the vanishing of tunneling.   
It is well known \cite{thirteen, sixteen} that the
appropriate space on which the continuous classical mechanics
is defined for the spin system is not
the unit sphere $S^2$ which is simply connected but the group manifold of $SO(3) = SU(2)/Z_2$
which is doubly connected.  The path integral quantization of the continuous
classical model on  the group manifold of
$SO(3)$ gives rise to two inequivalent sectors corresponding
to integer and half-integer spin.  The spin coherent-state representation leads to
the path integral on $S^2$.  A Wess--Zumino term arises
naturally to take care of the topological change of space.
Integrating out the momentum in the path--integral we obtain the configuration space
functional integral, i.e.
\begin{equation}
<{\bf n}_f|e^{-2i\hat{H}T}|{\bf n}_i> = e^{-i(\phi_f -\phi_i)s }{\cal K}(\phi_f, t_f=T; \phi_i, t_i=-T)
\end{equation}
where
\begin{equation}
{\cal K}(\phi_f, t_f;\phi_i, t_i) = \int \tilde{{\cal D}}{\phi}e^{i\int^{t_f}_{t_i} {\cal L}(\phi,\dot{\phi})dt}
\end{equation} 
The functional ${\cal K}$ is the Feynman propagator in configuration space with the
second order Lagrangian
\begin{displaymath}
{\cal L}= \frac {1}{2}m(\phi)\dot{\phi}^2 - V(\phi)
\end{displaymath}
and the measure 
\begin{displaymath}
\tilde {{\cal D}}{\phi} = \Pi^{M-1}_{k=1} \sqrt{\frac {m(\phi_k)}{2\pi i\epsilon}}d\phi_k
\end{displaymath}
The canonical momentum is defined as usual, i.e. $p = m(\phi)\dot{\phi}$.
The dynamical system is then reduced to that on the circle $S^1$.The  
potential is periodic with period $\pi $ and has an infinite number of
degenerate minima by successive $2\pi$ extension at $\Phi_n = n\pi$, $n$ being an integer.
Considering the periodic potential as a super-lattice with lattice constant
$\pi$, we can derive the energy spectrum in the tight-binding approximation.  
We let ${|m, \Phi_n>}$ be the set of low-lying eigenstates of the
zero order (i.e. oscillator-approximated) Hamiltonian $\hat{H}_0$ in the
$n$th well near the local minimum such that
\begin{equation}
\hat{H}_0|m,\Phi_n> = \epsilon_m|m,\Phi_n>
\end{equation}
and we define $|\psi>$ as the eigenstate of the full Hamiltonian with translation symmetry,
$\hat{H}(\phi) = \hat{H}(\phi + n\pi)$ and $\hat{H}|\psi> = E|\psi>$.
The state $|\psi>$ is then the (unnormalized) Bloch state
\begin{equation}
|\psi> = {\Sigma}_n e^{i\xi\Phi_n}|m, \Phi_n>
\end{equation}
Here $|\psi>$ and $|m, \Phi_n>$ denote the usual Dirac kets.  Eqs. (11) and (12) 
just define the ordinary quantum mechanical problem where $\xi$ is the
Bloch wave number.  Since the range of the azimuthal angle $\phi$ is $2\pi$,
or in other words the physical length of the superlattice is $2\pi$,
the periodic boundary condition
\begin{displaymath}
e^{i\xi\Phi_n} = e^{i\xi\Phi_{n+2}}
\end{displaymath}
implies that $\xi$ must be an integer.  The character of spin comes into the picture
only through the spin coherent-state representation which leads
to a topological phase of wave functions resulting directly from the
Wess--Zumino term.  
The energy ``band'' is seen to be given by
\begin{equation}
E = \epsilon_m - 2\triangle \epsilon \cos (s + \xi)\pi
\end{equation}
where the expression $\triangle \epsilon$ is given by
\begin{equation}
\triangle \epsilon =  - \int u^*_m(\phi, \Phi_n) \hat{H} u_m(\phi, \Phi_{n+1}) d\phi
\end{equation}
which is the usual overlap integral or simply the level shift due to
tunneling  through any one of the barriers
where $u_m(\phi, \Phi_n)$ denotes an eigenfunction of the Hamiltonian
$\hat{H_0}$ for the nth well.
The energy formula eq. (13) is only {\it formally} the same as that of the energy ``band'' of 
a one--dimensional lattice.  In our case there are only two levels for integer $s$
and there is only one level for half--integer $s$.   
The instanton method of calculation is nothing
but a procedure to evaluate the overlap integral analytically. If $s$ is half-integral
the energy ``band'' of eq.(13) shrinks to a single level which is doubly degenerate  \cite{seventeen} -- 
in agreement with Kramer's theorem and with numerical calculations \cite{six}.

However, the quenching of MQC is due to the vanishing of
$\cos (s+\xi)\pi$ and not to the vanishing of tunneling (i.e. $\triangle\epsilon = 0$)
as conjectured in ref.\cite{six}.

Since $\phi$ ranges from $0$ to $2\pi$, the Bloch waves
\begin{displaymath}
\psi(\phi) = \Sigma_n e^{i(s+\xi)\Phi_n} u_m(\phi, \Phi_n)
\end{displaymath}
(remembering that in the spin coherent-state representation the state $|\psi>$
becomes the wave function $<\phi|\psi> = \psi(\phi) e^{-is\phi}$)
then obey the periodic boundary condition for integer spin, and the antiperiodic
boundary condition for half-integer spin. Therefore the wave function of
the giant spin is a superposition of macroscopically distinct states
and changes sign under rotation of $2\pi$ for a half-integer spin s.
The ferromagnetic particle with half-integer spin $s$ may be looked at as a
macroscopic Fermi particle.  Up to this point we worked only
in real time.  The above results do not depend on Euclidean
paths inside the barrier in disagreement with the formalism in ref.\cite{six}. 

We now calculate the tunneling parameter, i.e. the level splitting $\triangle\epsilon$ and
hence the tunneling rate $\frac {2\triangle\epsilon}{\pi}$ which does not
vanish for half--integer spin $s$.
Passing to imaginary time by Wick rotation $\tau = it, \beta = iT$, the
amplitude for tunneling from the initial well, say that with $n=0$
(and $\phi_i = 0$), to the neighboring well with $n=1$ (and $\phi_f = \pi$)
and considering large $\beta$, the transition amplitude between the
corresponding coherent states is seen to be
\begin{eqnarray}
<{\bf n}(\pi)|e^{-2\beta\hat{H}}|{\bf n}(0)>
&=& <{\bf n}(\pi)|0, \Phi_1><0,\Phi_0|{\bf n}(0)> e^{-2\beta\epsilon_0}
\sinh (2\beta\triangle\epsilon)\nonumber\\
&=& e^{-i\pi s}{\cal K}_E(\phi_f=\pi, \beta; \phi_i = 0, -\beta)\nonumber\\  
\end{eqnarray}
where
\begin{displaymath}
{\cal K}_E = \int{\cal D}{\phi} e^{-S_E}
\end{displaymath}
is the Euclidean propagator with Euclidean action defined by
\begin{equation}
S_E = \int^\beta_{-\beta}{\cal L}_E d\tau, \;\ {\cal L}_E=\frac{1}{2}m(\phi)\dot{\phi}^2 + V(\phi)
\end{equation}
In eq.(15) $\epsilon_0$ is the ground state energy of $\hat{H}_0$. 
From now on $\dot{\phi} =\frac{d\phi}{d\tau}$ denotes the imaginary time derivative. The
phase in eqs. (6) and (9) for the case $\phi_f = \pi, \phi_i = 0$ is just the phase
relating to wave functions in the $0$th and $1$st wells in the coherent state
representation.

In the following the Euclidean propagator ${\cal K}_E$  is evaluated with the instanton method.
After evaluation we compare the result with eq.(15) to find the level splitting 
$\triangle\epsilon$.  The instanton configuration which minimizes the
Euclidean action $S_E$ is
\begin{equation}
\phi_c = \arcsin [\cosh^2\omega_0(\tau -\tau_0) - \lambda\sinh^2\omega_0(\tau -\tau_0)]^
{-\frac{1}{2}}
\end{equation}
with position $\tau_0$ and $\omega_0^2 = 4K_1K_2s^2$. 
The Euclidean action evaluated for the instanton trajectory eq.(17),
sometimes called the instanton mass, is
\begin{equation}
S_c = \int^\infty_{-\infty}m(\phi_c)\dot{\phi}^2_c d\tau = s\ln{ \frac{1+\sqrt\lambda}{1-\sqrt\lambda}}
\end{equation}
in agreement with refs.\cite {five, twelve}.  The functional integral ${\cal K}_E$ can be
evaluated with the stationary phase method by expanding $\phi$ about the instanton trajectory
$\phi_c$ such that $\phi = \phi_c + \eta$, where $\eta$ is the small fluctuation
with boundary conditions $\eta(\beta) = \eta(-\beta) = 0$.  Up to the one--loop
approximation we have
\begin{equation}
{\cal K}_E = e^{-S_c}I
\end{equation}
where
\begin{equation}
I = \int^{\eta(\beta)=0}_{\eta(-\beta)=0}{\cal D}\eta e^{-\delta S_E}
\end{equation}
is the fluctuation functional with the fluctuation action
\begin{equation}
\delta S_E=\int^{\beta}_{-\beta}\eta \hat{M}\eta d\tau
\end{equation}
where
\begin{equation}
\hat{M} = -\frac{1}{2}\frac{d}{d\tau}m(\phi_c)\frac{d}{d\tau} + \tilde{V}(\phi_c)
\end{equation}  
with
\begin{equation}
\tilde{V}(\phi_c) =\frac{1}{2}[-m^{\prime}(\phi_c)\ddot{\phi}_c -
\frac {1}{2}m^{\prime\prime}(\phi_c)\dot{\phi}_c^2 + V^{\prime\prime}(\phi_c)]
\end{equation}
Here $\hat{M}(\phi_c)$ is the operator of the second variation of the action
and $m^{\prime}(\phi_c) = {\frac{\partial m(\phi)}{\partial\phi}}|_{\phi=\phi_c}$. As in the usual method
of evaluating
the fluctuation integral $I$, we expand the fluctuation variable $\eta$ in terms of the
eigenmodes of $\hat{M}$ and set $\eta = \Sigma_n C_n\psi_n$, where $\psi_n$ 
denotes the $n$th eigenfunction of $\hat{M}$, and express the result of the integration  as an inverse
square root of the determinant of $\hat{M}$.  In view of the time translation symmetry
of the equation of motion, the functional integral ${\cal K}_E$ is not well defined when expanded
about the classical solutions.  The translational symmetry results
in zero eigenmodes of the second variation operator $\hat{M}$ of the action (which in the present
case, of course, has only one).  This problem can be cured by the Faddeev--Popov
procedure \cite{eightteen} or in a more systematic way with the help of the BRST transformation
\cite{nineteen}. Then the zero--mode functional integral is converted into an integral
over the collective coordinate, i.e. the instanton position $\tau_0$, and leads to
$2\beta$ and a Faddeev--Popov determinant $\sqrt{D}$ with
\begin{equation}
D = \int^{\infty}_{-\infty}{\dot{\phi}_c}^2d\tau = \omega_0\Bigg(1+\frac{1-\lambda}{2\sqrt\lambda}\ln
{\frac{1+\sqrt\lambda}{1-\sqrt\lambda}}\Bigg)
\end{equation}

The complicated determinant of $\hat{M}$ and the related
Jacobian are avoided by the direct integration involved in the transformation of the 
well--known shift method \cite{nineteen, twenty}, i.e.
\begin{equation}
\eta = y + \dot{\phi}_c\int^{\tau}_{\tau_i} \frac{{\ddot{\phi}}_c(\tau^{\prime})}
{\dot{\phi}^2_c(\tau^{\prime})} y(\tau^{\prime})d\tau^{\prime}
\end{equation}
where the derivative of the instanton solution, i.e.$\dot{\phi_c}$, is simply the zero
eigenmode of $\hat{M}$.
The resulting fluctuation integral is
\begin{equation}
I = \frac{1}{\sqrt{2\pi}}[\dot{\phi}_c(\beta)\dot{\phi}_c(-\beta)\int^{\beta}_{-\beta}
\frac {d\tau}{\dot{\phi}^2_c(\tau)m(\phi_c(\tau))}]^{-\frac{1}{2}}
\end{equation}
Following the procedure of ref.\cite{nineteen} or earlier work of ref.\cite{twenty}
the one instanton contribution to the propagator in the one-loop
approximation is calculated to be
\begin{equation}
{\cal K}^{(1)}_E = 2\beta\frac {4\sqrt 3}{\pi} (1-\lambda)^{-\frac{1}{2}}s^2 K_2
e^{- \omega_0\beta}e^{-S_c}
\end{equation}
To obtain the desired result proportional to $\sinh (2\beta\triangle\epsilon)$ in eq.(15),
the contributions
of the infinite number of instanton and antiinstanton pairs to the one instanton
contribution have to be taken into account.  Interactions among instantons and
antiinstantons are neglected in the dilute instanton--gas approximation.
The contribution of one instanton plus n such pairs to the propagator is obtained
with the help of the group property of the Feynman path--integral, i.e.
\begin{equation}
{\cal K}^{(2n+1)}_E =\frac {(2\beta)^{2n+1}}{(2n+1)!}(\frac{s}{2\pi})^{\frac{1}{2}}
\lambda^{\frac {1}{4}}[2^{\frac {5}{2}}\{\frac {3K_1K_2}{(1-\lambda )\pi}\}^{\frac{1}{2}}
\lambda^{\frac{1}{4}}s^{\frac{3}{2}}]^{2n+1} e^{-(2n+1)S_c}e^{-\omega_0\beta}
\end{equation}
Summing over all contributions ${\cal K}^{(2n+1)}_E$, the final result of the propagator
is found to be
\begin{equation}
{\cal K}_E = \lambda^{\frac{1}{4}}(\frac{s}{2\pi})^{\frac{1}{2}}e^{-\beta\omega_0}\sinh[2\beta.
2^{\frac {5}{2}}\{\frac {3K_1K_2}{(1-\lambda )\pi}\}^{\frac {1}{2}}\lambda^{\frac{1}{4}}
s^{\frac{3}{2}}e^{-S_c}]
\end{equation}
Comparing with eq.(15) the level splitting is seen to be
\begin{equation}
\triangle\epsilon = 2^{\frac{5}{2}}\{\frac{3K_1K_2}{(1-\lambda)\pi}\}^{\frac{1}{2}}
\lambda^{\frac {1}{4}}s^{\frac {3}{2}}e^{-s\ln\frac{1+\sqrt{\lambda}}{1-\sqrt{\lambda}}}
\end{equation}
The product of the values of the wave functions at neighboring potential minima is
\begin{equation}
<{\bf n}(\pi), 0|0, \Phi_1><0, \Phi_0|{\bf n}(0)> = (\frac {s}{2\pi})^{\frac {1}{2}}
\lambda^{\frac {1}{4}}e^{-i\pi s}
\end{equation}
and the ground state energy of $\hat{H}_0$ is identified as $\epsilon_0 = \frac {1}{2}\omega_0$.
Expanding $\triangle\epsilon$ in the region of small values of $\lambda$, the level
splitting is
\begin{equation}
\triangle\epsilon = 2^{\frac {5}{2}}\{\frac {3K_1K_2}{\pi}\}^{\frac {1}{2}}
\lambda^{\frac{1}{4}}s^{\frac{3}{2}}
e^{-2s\sqrt{\lambda}}
\end{equation}
Since $s$ is a large number ($500$ to $5000$ as cited in the literature \cite{three}),
the level splitting is suppressed in leading order by the factor
\begin{displaymath}
e^{-S_c}\simeq e^{-2s\sqrt{\lambda}}
\end{displaymath}
For $\triangle\epsilon$ to be observable, plausible values of $\lambda$
are $<10^{-4}$ as discussed e.g. in ref.\cite{three}.
The quantum correction proportional to $e^{\frac{3}{2}\ln s}$ cannot increase the 
tunneling probability significantly, but may be of importance for small values of $\lambda$.
We remark that one of the crucial points of the effective Hamiltonian as well as
the Lagrangian for a quantum spin system is the position dependent mass which
requires great care in the evaluation of the path integral.  The result is not simply
the factor $\sqrt{S_c}$ as in the usual case of constant mass.

If in the case of the potential with degenerate minima only one of the
potential minima is considered as a perturbative vacuum; however, a fictitious imaginary
energy can be calculated by consideration of possible
back and forth tunneling in the barrier.  Therefore there exists a relation between the 
level splitting and this imaginary energy, and has been referred to 
as the Bogomolny--Fateyev relation \cite{ten, twentytwo, twentythree}
\begin{displaymath}
{\it Im}E = 2\pi(\triangle\epsilon)^2
\end{displaymath}
in the context of
large order perturbation theory (in \cite{ten, twentytwo, twentythree} the factor
$2\pi$ arises with $\omega_0=1$).  Recently the relation has been rederived
and generalized to low--lying levels with the instanton method by regarding the
instanton--antiinstanton pair as a bounce--like configuration \cite{ten, twentythree}.

In the literature \cite{six} the quenching of the energy ``band'' structure for half--integer $s$
(here, of course, we have only formally a ``band'')
has been interpreted as due to the quenching of tunneling.  The level splitting or
overlap integral is said to vanish due to an interference of the Euclidean paths
with the topological phase in eqs.(6) and (9).  The destructive interference, however, does not
occur for closed loops because the interference in this case would be proportional to
$|e^{i2\pi s}+ e^{-i2\pi s}|$ which does not vanish for half--integer values of $s$.
Thus there is an apparent contradiction with the Bogomolny--Fateyev relation from the viewpoint
of quenching of tunneling: the level splitting is said to vanish but its
theoretical partner, namely the imaginary part which is
calculated from the bounce (a closed Euclidean path) does not.
Therefore we suspect a misinterpretation in ref.\cite {six} as we indicated in our earlier remark.
There is no
contradiction in our framework.  The level splitting or overlap integral does
not vanish, which is, as it ought to be.

In conclusion we can say: In a theoretical analysis of MQC of ferromagnetic particles, 
the periodic nature of the Hamiltonian must be taken into account. The 
quenching of MQC results from interference of Bloch waves which  obey
an antiperiodic boundary condition. 
The tunneling cannot be suppressed for half--integer spin $s$ in
a consistent instanton method.  
 The ferromagnetic particle with half--integer spin
$s$ may be looked at as a macroscopic Fermi particle configuration.  The correction
of quantum fluctuations with the position dependent mass
to the level splitting for MQC, here first reported \cite{twentyfour}, does
not enhance the tunneling probability significantly, as expected.
Nevertheless the contribution of quantum fluctuations increases the tunneling
probability along with $s$, which is proportional to the volume of the
giant particle. In the above eqs. (30) and (32) provide explicit formulas
of the level splitting suitable for quantitative evaluation. 
Enhancement of MQC may take place theoretically if tunneling from an excited state
is considered, even for low--lying levels.  This investigation will be reported 
elsewhere.

\subsection*{Acknowledgement}

J.--Q.Liang  acknowledges support of the Deutsche Forschungsgemeinschaft and J.--G.
Zhou support of the A. von Humboldt Foundation. J.--Q.Liang also thanks Professor F.C.
Pu for helpful discussions and suggestions.

\end{document}